\documentclass[runningheads,citeauthoryear]{apinv}
\usepackage{epsfig,cite,graphics}
\usepackage[T2A]{fontenc}
\usepackage[cp1251]{inputenc}

\newcommand{\ltsima} {$\; \buildrel < \over \sim \;$} 
\newcommand{\simlt}  {\lower.5ex\hbox{\ltsima}}            % < over MMM 
\newcommand{\gtsima} {$\; \buildrel > \over \sim \;$} 
\newcommand{\simgt}  {\lower.5ex\hbox{\gtsima}}            % > over MMM 

\newcommand{\msun}{$M_\odot$}
\newcommand{\rsun}{$R_\odot$}
\newcommand{\lsun}{$L_\odot$}

\def\kms{km~s$^{-1}$}   
\def\e{$\pm$}

\begin{document}

\title{Emission line variability in the spectrum of V417~Centauri
\thanks{based on data from ESO (program 073.D-0724) and AAVSO} }
\titlerunning{Emission line variability in the spectrum of V417~Centauri}
\author{K. A. Stoyanov$^{1}$\thanks{e-mail: kstoyanov@astro.bas.bg}, R. K. Zamanov$^{1}$,
M. F. Bode$^{2}$, J. Pritchard$^{3}$,  N. A. Tomov$^{1}$, A. Gomboc$^{4,5}$, K. Koleva$^{1}$ }
\authorrunning{K. Stoyanov, R. Zamanov, M. Bode et al.}
\tocauthor{K. Stoyanov, R. Zamanov, M. Bode et al.} 
% Command tocautor{} is used by the Latex to give author names 
% to the Contents of the volume (automatically generated)
\institute{Institute of Astronomy and National Astronomical Observatory, Bulgarian Academy of Sciences,  
      72 Tsarighradsko Shousse Blvd.,  1784 Sofia, Bulgaria 
	\and Astrophysics Research Institute, Liverpool John Moores University, 
IC2 Liverpool Science Park, 146 Brownlow Hill, Liverpool L3 5RF, UK 
	\and European Southern Observatory, Karl-Schwarzschild-Str. 2, D-85748 Garching bei M\"{u}nchen, Germany
	\and Faculty of Mathematics and Physics, University of Ljubljana, Jadranska 19, 1000 Ljubljana, Slovenia 
	\and Centre of Excellence SPACE-SI, A\v{s}ker\v{c}eva cesta 12, SI-1000, Ljubljana, Slovenia  }
\papertype{Research report. Accepted on xx.xx.xxxx}	
% Papertype can be "Research report", "Review", "Invited lecture", "Conference talk", 
% "Conference poster", "Lecture at scientific seminar", "Summary of dissertation",  etc.
\maketitle

\begin{abstract}
We report high resolution ($ \lambda / \Delta \lambda  \sim 48000 $) spectral observations of the yellow symbiotic star 
V417~Cen obtained in 2004. We find that the equivalent widths of the emission lines decreased, while the brightness increased.
The FWHMs and wavelengths of the emission lines do not change.

We estimated the interstellar extinction towards  V417~Cen as  $E_{B-V}$ = 0.95 $\pm0.10$, using the KI interstellar line. 

Using the [O~III] lines, we obtain a rough estimation of the density and the temperature in the forbidden lines region  
$\log N_e \approx 4.5 \pm 0.5$ and $T_e=100000 \pm 25000$~K.
Tidal interaction in this binary is also discussed. 

\end{abstract}

\section{Introduction}
Symbiotic stars (SSs) are thought to comprise a compact object (usually a white dwarf) accreting from a cool giant.
They offer a laboratory in which to study such processes as 
(1) mass loss from cool giants and the formation of nebulae, 
(2) accretion onto compact objects, radiative transfer in gaseous nebulae, 
(3) jets and outflows   (i.e. Corradi, Mikolajewska \& Mahoney 2003).\\

V417~Centauri (HV~6516, Hen~3-977) is a poorly studied D'-type (yellow) symbiotic system 
surrounded by a faint asymmetric nebula.  The symbiotic nature of the object was proposed by 
Steiner, Cieslinski \& Jablonski (1988).
The cool component is a G2~Ib-II star, with $ \log (L/L_\odot) =3.5$ and $T_{eff}=5000$~K 
(van Winckel et al. 1994). Pereira,  Cunha \& Smith  (2003) found  atmospheric parameters
 $T_{eff} = 6000$,  $\log g =3.0$, and spectral type F9 III/IV.

The binary period  is not defined. 
Van Winckel et al. (1994) found a 245.68~day periodicity
with an amplitude of 0.5 mag using Harvard and Sonneberg plates. 
Gromadzki et al. (2011) using optical photometric observations 
covering  20 years detected strong long term modulation with a period 
of  about 1700 days and amplitude about 1.5 mag in V-band, in addition to variations 
with shorter time-scales and lower amplitudes. However, the long period seems to be non-coherent and the
nature of light variations and the length of the orbital period remain unknown.

Here we discuss the emission line variability of V417~Cen in 2004.

\section{Observations}
We secured 8 spectra in 2004 using the ESO, La Silla, 2.2m telescope and the FEROS spectrograph. FEROS is a
fibre-fed echelle spectrograph, providing a resolution of $\lambda$/$\Delta\lambda$=48000, wide wavelength
coverage from about 4000~\AA\ to 8000~\AA\ in one exposure and a high detector efficiency (Kaufer et al. 1999).
The log of observations is given in Table~\ref{tab1}, and the measured emission line parameters in Table~\ref{tab2}.
The typical errors in the equivalent widths  are $\pm$10\% for the strong lines (EW$>$3~\AA)
and  $\pm$20\% for the weaker lines (EW  $ \simlt$ 3~\AA). Errors for the fluxes are typically 10\% and 20\%;  
FWHM around 0.02 - 0.10 \AA, and for wavelengths about 0.1 ~\AA.  
On Fig.~\ref{spec} are plotted spectra, obtained during three nights.
The spectrum from June 3, 2004 is almost  identical to the spectrum in July 1993 (see Fig.3 of Van Winckel et al. 1994). 
\section{Interstellar extinction towards V417~Cen}

%%%--------------------------------------------------------------------------------
 \begin{figure} 
 \vspace{8.0cm}  
  \includegraphics{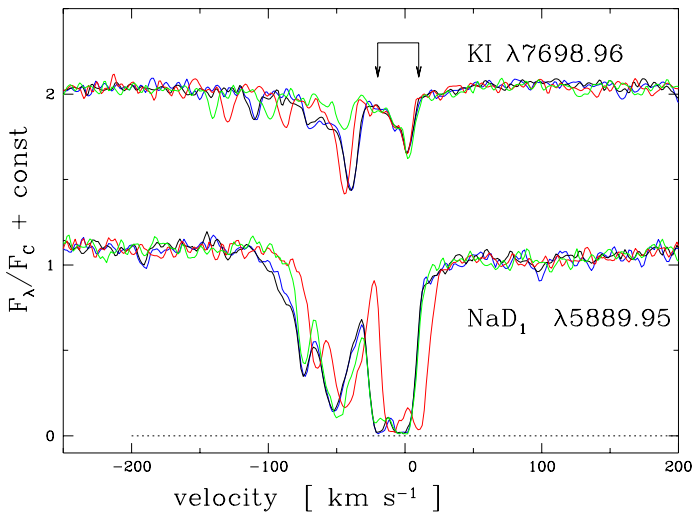}   
  \caption[]{The lines  KI~$\lambda$7699.96 (upper) and  NaD~$\lambda$5889.95 (lower plot). 
  The spectra are normalized to the local continuum. The blue line represents sp1, the black line - sp3, 
  the red line - sp5, and the green line - sp8. The arrows indicate the  non-variable part of the
  KI line, on which basis we estimate $E_{B-V} = 0.95  \pm 0.10$ (see the text).}		    
\label{NaD}     
\end{figure}	    

The NaD$_1$ and NaD$_2$ lines as well as the KI~$\lambda$7699 line are visible in our spectra
and are plotted in Fig.\ref{NaD}. A clear variability is visible indicating that 
they  have both stellar and interstellar origin.  NaD lines are saturated and inappropriate for 
calculation of the interstellar extinction. We do detect a non-variable part in KI~$\lambda$7699
which has  EW (KI~$\lambda$7698.965) $ \approx $ 0.24 \AA . 
%  (even some additional absorption   EW (KI 7698.965) $\le$ 0.30 \AA )
Following the calibration of  Munari \& Zwitter (1997) this corresponds to $E_{B-V}$ = 0.95 $\pm0.10$.
This value is  similar  to  the previous estimates of
$E_{B-V} = 1.15$ (Van Winckel et al. 1994, obtained on the basis of HI line ratios) 
and $E_{B-V} \sim 0.7$ (Cieslinski, Elizalde \& Steiner 1994 from the photometric colours). 

%%%--------------------------------------------------------------------------------
 \begin{figure*} 
  \vspace{10.5 cm}  
  \includegraphics{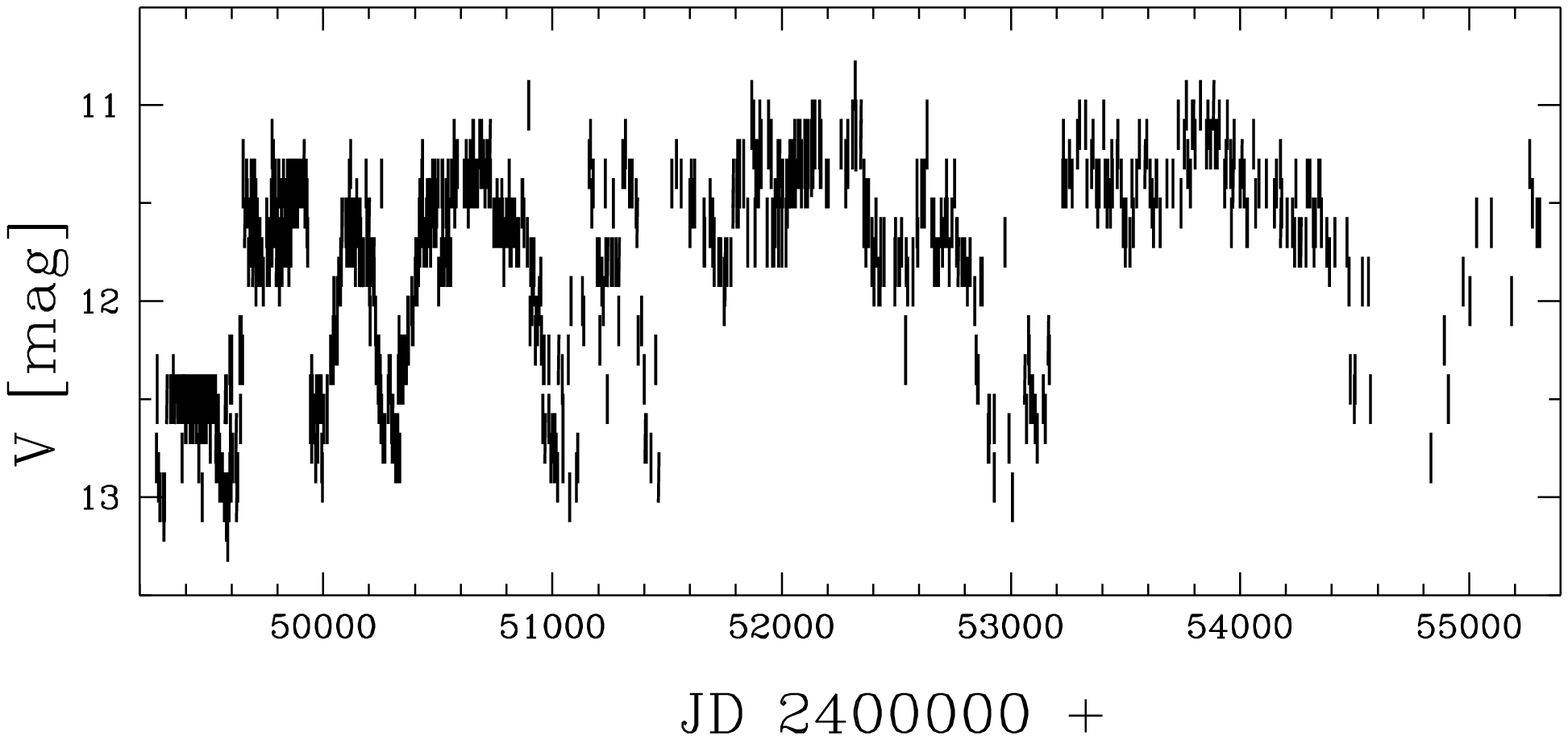}   
  \includegraphics{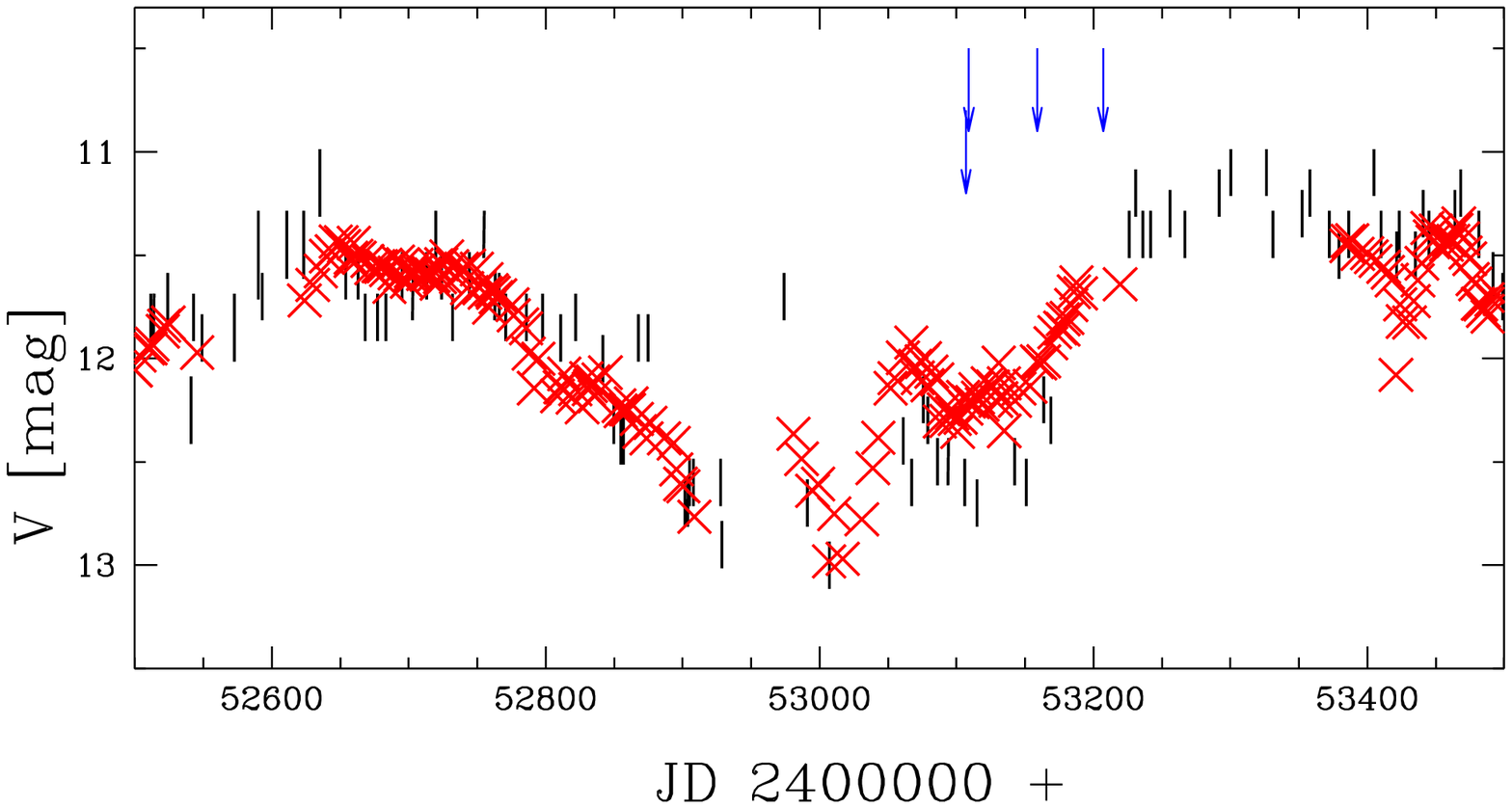}   
  \caption[]{Brightness variability of V417~Cen from American Association of Variable Star Observers (AAVSO) and ASAS observations. The (black) lines indicates the data from AAVSO and the (red) crosses - the data from ASAS. The arrows indicate the time of our
             spectroscopic observations.}
  \label{AAVSO}
  \vspace{11.7cm}  	     	
  \includegraphics{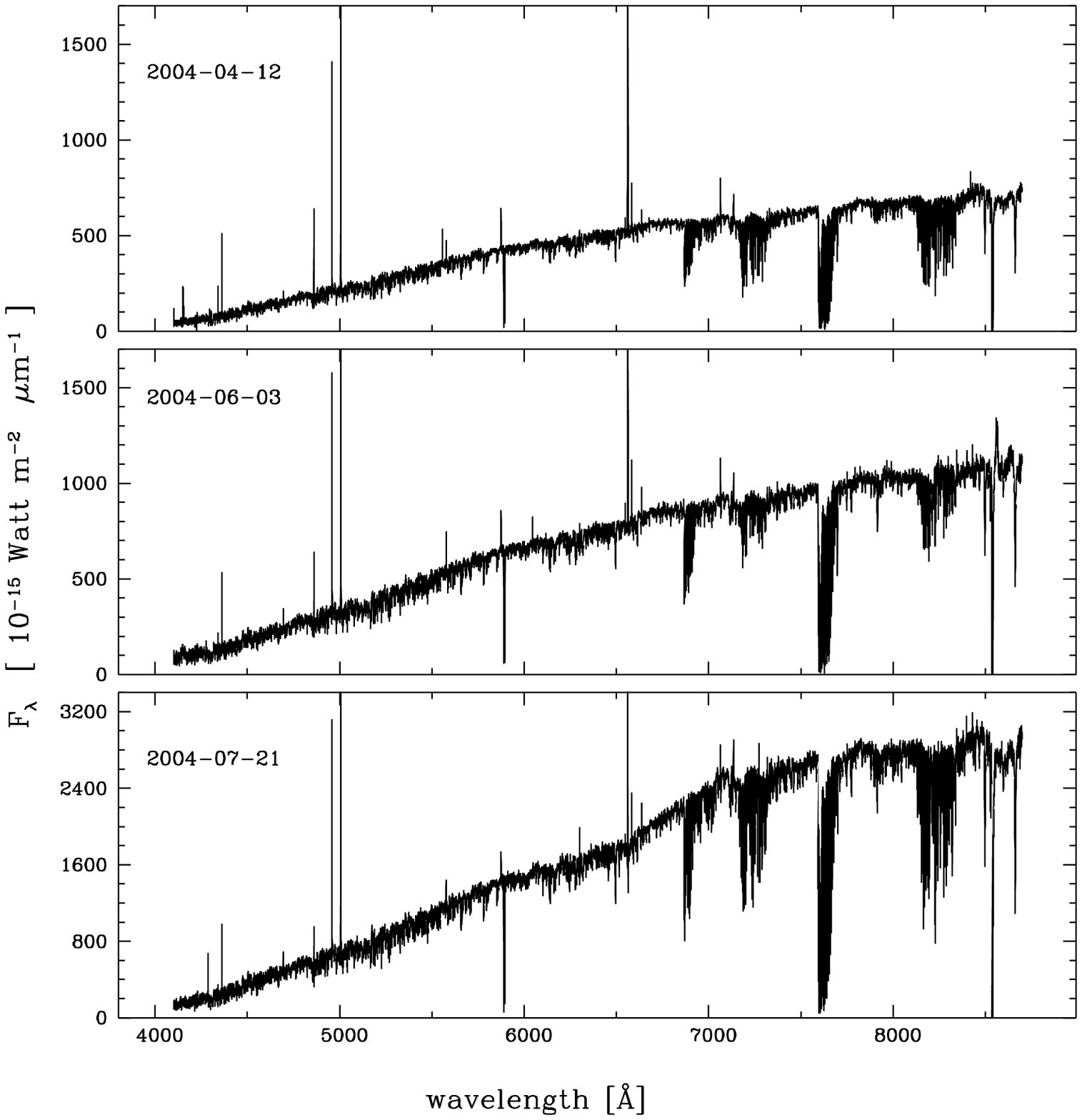}   
  \caption[]{The spectra of V417~Cen as obtained on  April 12, June 3, and July 21 (from top to bottom panel), 
  degraded to  0.3 \AA /pix. }		    
  \label{spec}     
\end{figure*}	    	    
%%%----------------------------------------------------------------------------------

%%%--------------------------------------------------------------------------------
 \begin{figure*} 
 \vspace{17.0cm}  
    \includegraphics{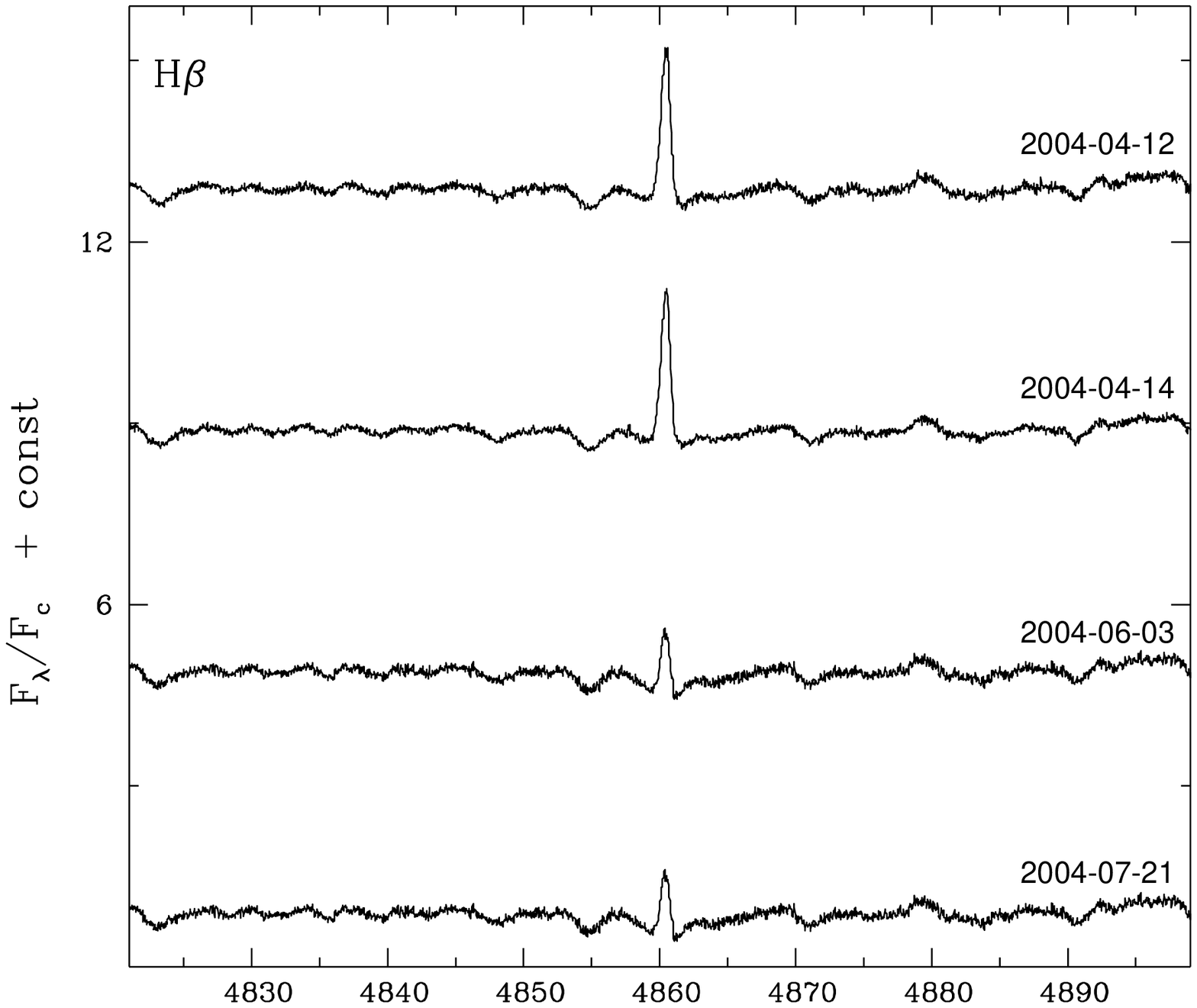}   
    \includegraphics{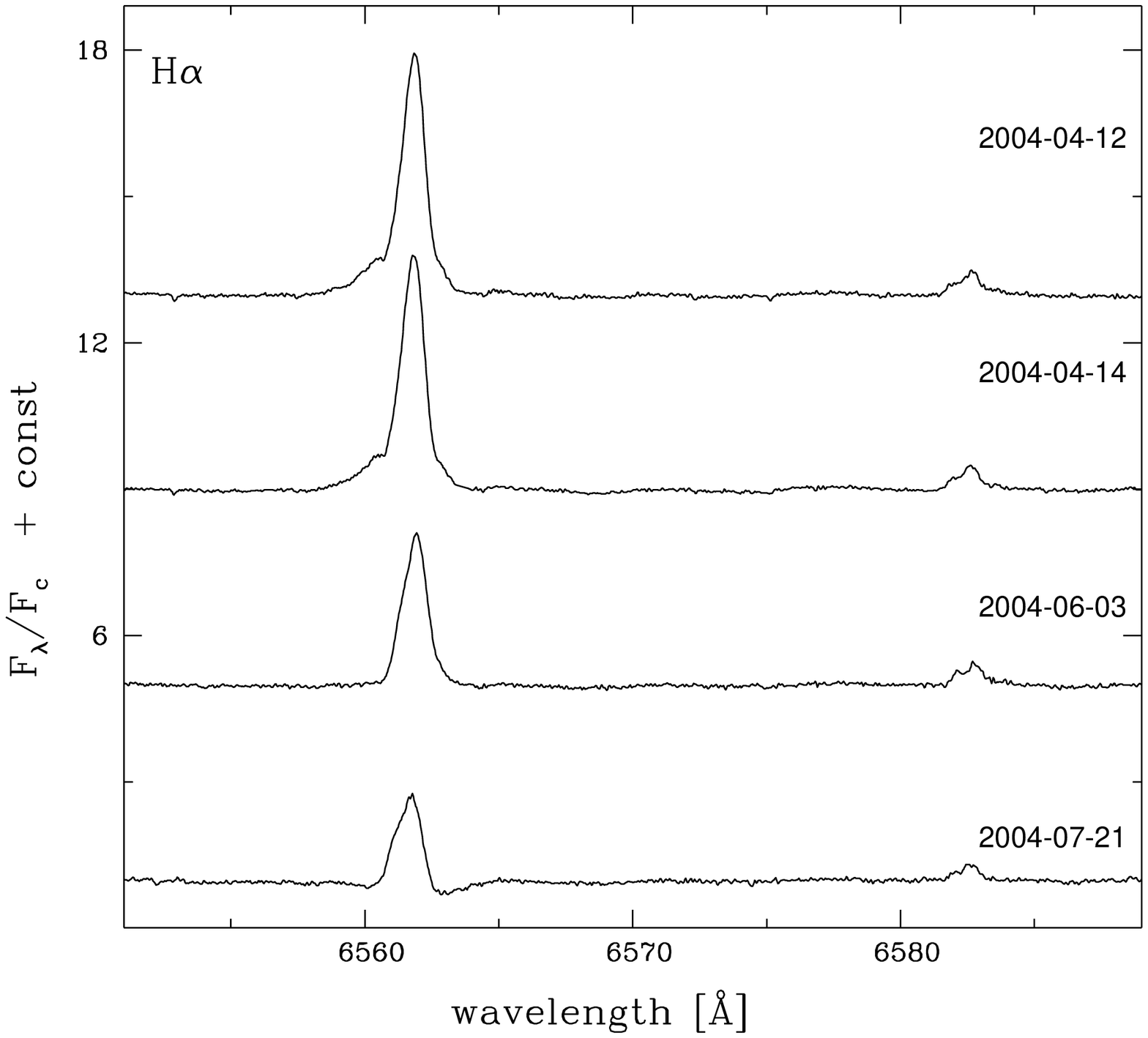}   
    \includegraphics{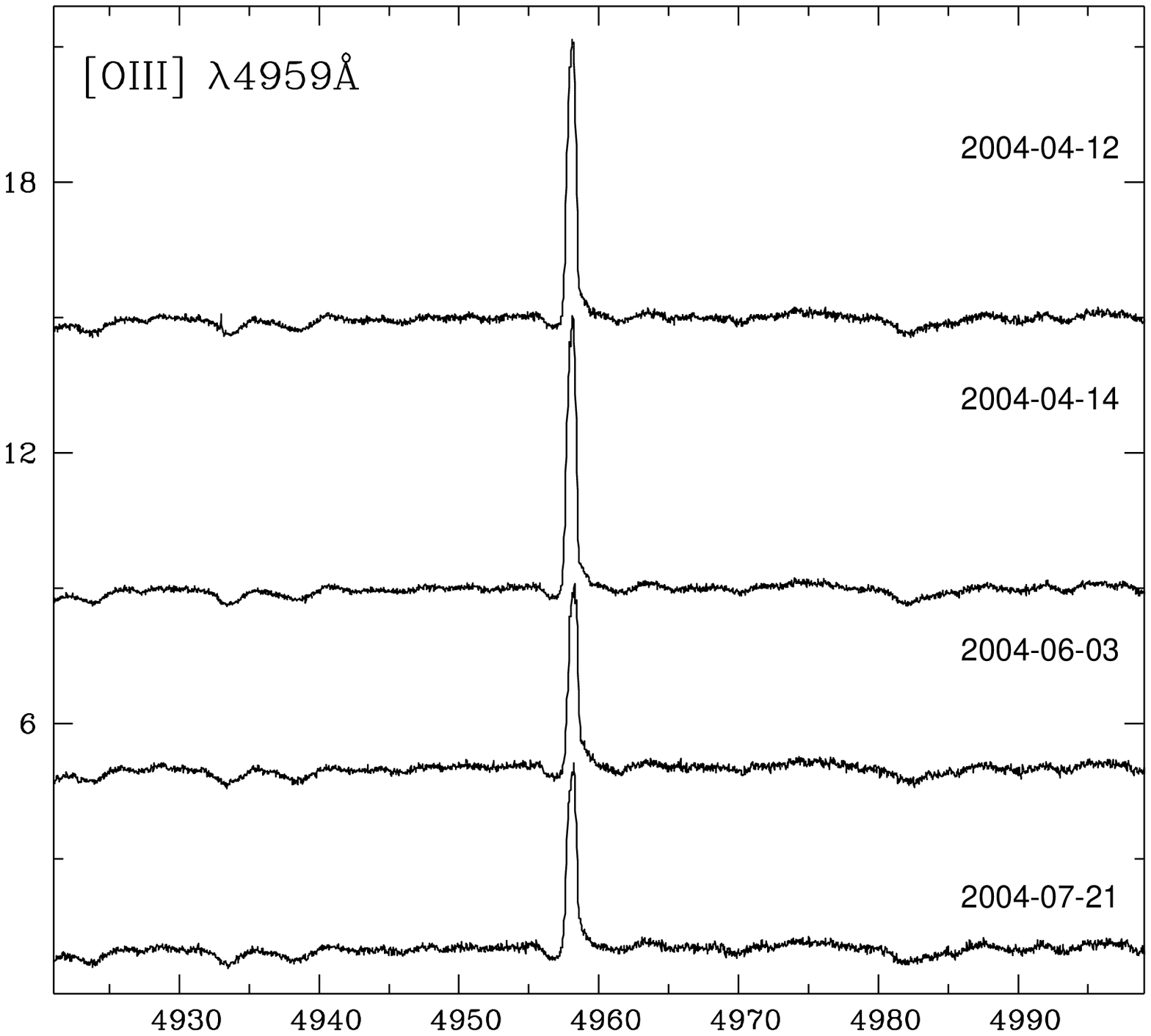}	
    \includegraphics{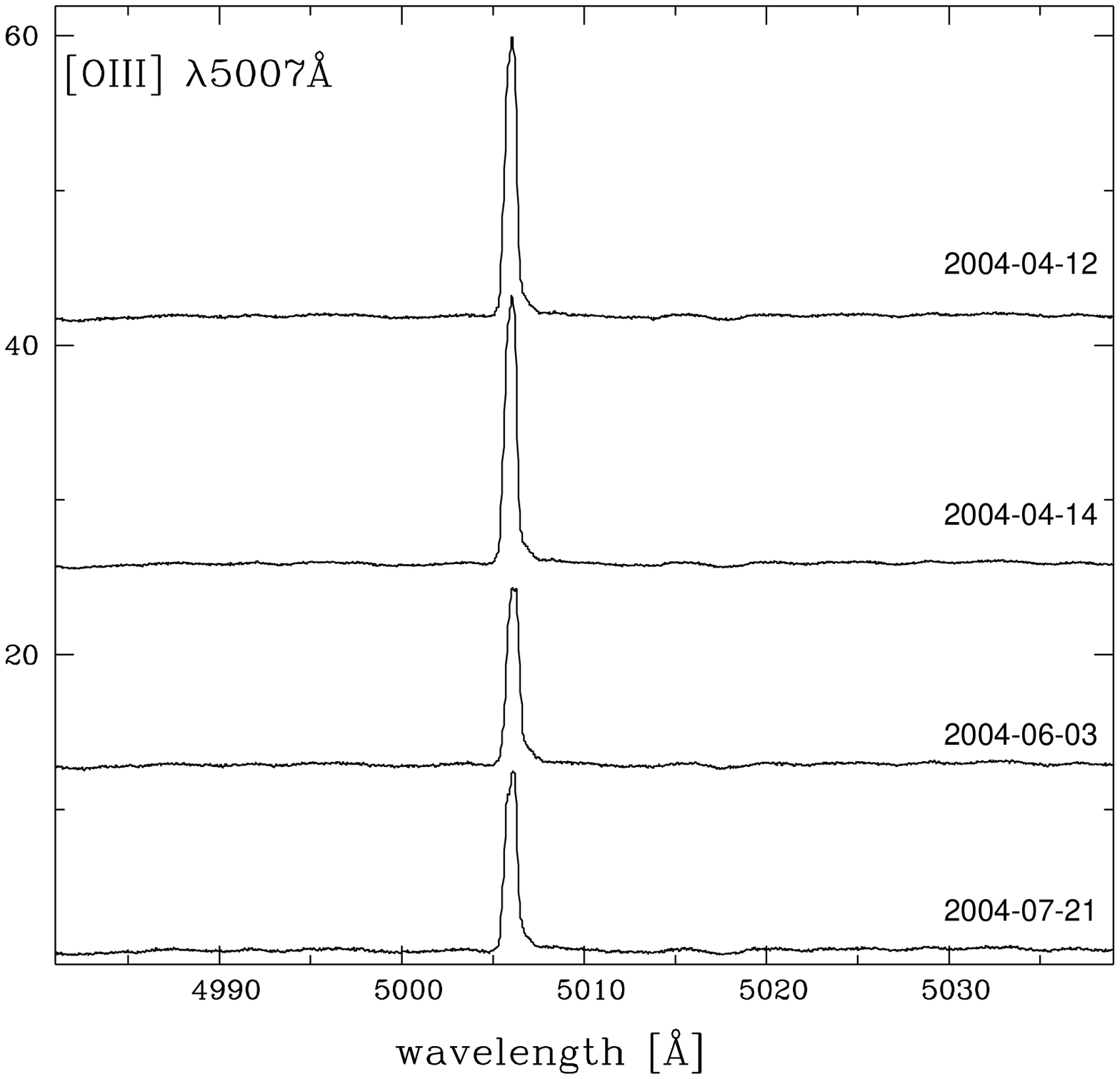}	  
  \caption[]{The H$\beta$, H$\alpha$ (left), 
            [OIII]$\lambda$4959 and [OIII]$\lambda$5007 (right) emission lines of V417~Cen. 
            All spectra are are normalized to the local continuum.}		   
  \label{halpha}     
\end{figure*}	    
%%%----------------------------------------------------------------------------------

%%%------------------------------------------------------------
\begin{table} %[!hb]
\footnotesize
\caption{Log of observations of V417~Cen. In the table are given the number of the spectrum, 
date of observations, Julian Day, and exposure time.}
\begin{center}
\begin{tabular}{llrrrrcrrccccrlcr}
\hline

N:    & Date         &   JD      & exptime &  &  & \\
      & yyyy-mm-dd   & 2400000+  &  [sec]  &  &  & \\
\hline
sp1     & 2004-04-12 & 53107.254 & 600   &  &  & \\
sp2     & 2004-04-12 & 53107.262 & 600   &  &  & \\
sp3     & 2004-04-14 & 53109.273 & 600   &  &  & \\
sp4     & 2004-04-14 & 53109.280 & 600   &  &  & \\ 
sp5     & 2004-06-03 & 53159.188 & 600   &  &  & \\
sp6     & 2004-06-03 & 53159.195 & 600   &  &  & \\
sp7     & 2004-07-21 & 53207.006 & 600   &  &  & \\
sp8     & 2004-07-21 & 53207.013 & 600   &  &  & \\
%----------------------------------------------
% \caption{Emission lines  of V417~Cen. In the table are given the fluxes  in  [erg cm$^{-2}$ s$^{-1}$].  
% The typical errors are  10-30 \%. }
\\
\hline
\\
\end{tabular}
\end{center}
\label{tab1}
\end{table}
%%%--------------------------------------------------------------------------------------

%%%------------------------------------------------------------
\begin{table*} %[!hb]
\footnotesize
\caption{Optical emission line parameters. 
The typical errors in the equivalent widths and fluxes are $\pm$10\% for the strong lines (EW$>$3~\AA)
and  $\pm$20\% for the weaker lines (EW  $ \simlt$ 3~\AA).  
%   \hskip 1.0cm {\bf /work/V417Cen/st2.sp.V417Cen}  .   
}
\begin{center}
\begin{tabular}{lrrrrrrrcrlcr}
\hline
spec No:          & sp1    & sp2  &sp3    & sp4     &sp5    & sp6  &  sp7     & sp8   &    \\
        	  & \multicolumn{2}{c}{JD53107} & \multicolumn{2}{c}{JD53109} & \multicolumn{2}{c}{ JD53159 }  & \multicolumn{2}{c}{JD53207}   &    \\
\hline
\multicolumn{3}{c}{Equivalent Widths [\AA]} & \\
4101	H$\delta$ &  1.04  &  0.87 &  1.24  &  1.14 & 0.33 &  0.32 & 0.07  &  0.14 & \\	 
4340	H$\gamma$ &  2.57  &  2.20 &  2.78  &  2.19 & 1.24 &  1.26 & 0.96  &  0.99 & \\  
4363	[O III]   &  2.95  &  2.44 &  2.45  &  2.33 & 1.52 &  1.41 & 1.47  &  1.55 & \\
4861	H$\beta$  &  2.88  &  3.07 &  2.67  &  2.75 & 1.64 &  1.60 & 1.56  &  1.39 & \\  
4959	[O III]   &  4.59  &  4.76 &  4.37  &  4.78 & 3.33 &  3.43 & 3.23  &  3.01 & \\  
5007	[O III]	  & 13.89  & 13.68 & 13.97  & 13.08 & 9.85 &  9.72 & 9.53  &  9.05 & \\		   
5876	He I	  &  0.48  &  0.49 &  0.52  &  0.51 & 0.32 &  0.25 & 0.18  &  0.20 & \\				      
6548	[N II]	  &  0.11  &  0.11 &  0.12  &  0.12 & 0.16 &  0.10 & 0.07  &  0.06 & \\
6563	H$\alpha$ &  6.53  &  6.37 &  6.10  &  6.16 & 3.55 &  3.48 & 3.04  &  2.84 & \\  
6584	[N II]	  &  0.54  &  0.46 &  0.48  &  0.44 & 0.43 &  0.34 & 0.27  &  0.30 & \\  
7065	He I	  &  0.47  &  0.54 &  0.57  &  0.55 & 0.34 &  0.21 & 0.19  &  0.16 & \\
7135	[Ar III]  &  0.27  &  0.25 &  0.30  &  0.31 & 0.24 &  0.24 & 0.18  &  0.18 & \\				   
\hline	
\multicolumn{2}{l}{ FWHM [\AA] } &  \\	 
4101	H$\delta$ & 0.78\e0.05 & 0.84\e0.05 & 0.90\e0.05 & 0.65\e0.06 & 0.8\e0.4:  & 0.6\e0.3:  & ---        & 1.0\e0.4:  &   \\    
4340	H$\gamma$ & 0.67\e0.01 & 0.74\e0.03 & 0.75\e0.02 & 0.75\e0.02 & 0.73\e0.02 & 0.81\e0.02 & 0.65\e0.01 & 0.73\e0.05 &   \\  
4363	[O III]   & 0.51\e0.01 & 0.50\e0.01 & 0.51\e0.01 & 0.52\e0.01 & 0.56\e0.02 & 0.51\e0.02 & 0.53\e0.01 & 0.52\e0.03 &   \\
4861	H$\beta$  & 0.80\e0.01 & 0.82\e0.01 & 0.80\e0.01 & 0.81\e0.01 & 0.77\e0.02 & 0.78\e0.02 & 0.53\e0.01 & 0.78\e0.01 &   \\  
4959	[O III]   & 0.69\e0.01 & 0.69\e0.01 & 0.70\e0.01 & 0.70\e0.01 & 0.73\e0.02 & 0.76\e0.01 & 0.74\e0.01 & 0.74\e0.02 &   \\  
5007	[O III]	  & 0.70\e0.01 & 0.69\e0.01 & 0.70\e0.01 & 0.72\e0.01 & 0.75\e0.01 & 0.75\e0.01 & 0.75\e0.01 & 0.74\e0.01 &   \\ 		      
5876	He I	  & 0.96\e0.02 & 0.94\e0.02 & 1.05\e0.07 & 1.01\e0.03 & 1.07\e0.04 & 0.96\e0.02 & 0.91\e0.04 & 1.27\e0.04 &   \\ 				  
6548	[N II]	  & 0.70\e0.08 & 0.86\e0.05 & 0.88\e0.10 & 1.12\e0.10 & 1.30\e0.15 & 0.95\e0.10 & 1.0\e0.2:  & 1.3 \e0.3: &   \\
6563	H$\alpha$ & 1.15\e0.01 & 1.10\e0.02 & 1.11\e0.01 & 1.09\e0.01 & 1.11\e0.01 & 1.11\e0.02 & 1.16\e0.05 & 1.16\e0.01 &   \\  
6584	[N II]	  & 1.10\e0.10 & 1.14\e0.04 & 1.11\e0.02 & 1.09\e0.10 & 1.17\e0.05 & 1.12\e0.05 & 0.94\e0.03 & 1.08\e0.04 &   \\  
7065	He I	  & 1.43\e0.05 & 1.45\e0.07 & 1.49\e0.03 & 1.56\e0.04 & 1.10\e0.10 & 1.24\e0.02 & 1.25\e0.03 & 1.39\e0.10 &   \\
7135	[Ar III]  & 1.28\e0.08 & 1.27\e0.08 & 1.14\e0.02 & 1.16\e0.02 & 1.06\e0.10 & 1.17\e0.04 & 1.28\e0.06 & 1.38\e0.04 &   \\  
\hline
\multicolumn{4}{l}{Optical emission line fluxes [ erg~cm$^{-2}$~s$^{-1}$ ] }  &  \\
4101	H$\delta$ &  9.39E-15  &     6.39E-15	&     7.73E-15  &	7.07E-15  &	  1.73E-15   &	   3.06E-15  &	   6.52E-16   &    2.70E-15 & \\	   
4340	H$\gamma$ &  1.82E-14  &     1.76E-14	&     2.06E-14  &	1.92E-14  &	  1.14E-14   &	   1.39E-14  &	   1.13E-14   &    1.28E-14 & \\ 
4363	[O III]   &  3.40E-14  &     3.31E-14	&     3.21E-14  &	3.14E-14  &	  2.66E-14   &	   2.46E-14  &	   3.84E-14   &    3.92E-14 & \\
4861	H$\beta$  &  6.00E-14  &     6.12E-14	&     6.00E-14  &	5.74E-14  &	  4.06E-14   &	   4.44E-14  &	   3.89E-14   &    3.68E-14 & \\  
4959	[O III]   &  1.30E-13  &     1.37E-13	&     1.36E-13  &	1.34E-13  &	  1.11E-13   &	   1.23E-13  &	   1.55E-13   &    1.53E-13 & \\  
5007	[O III]	  &  4.20E-13  &     4.33E-13	&     4.23E-13  &	4.09E-13  &	  3.59E-13   &	   3.63E-13  &	   4.77E-13   &    4.64E-13 & \\	
5876	He I	  &  2.86E-14  &     2.67E-14	&     3.21E-14  &	3.20E-14  &	  2.16E-14   &	   2.41E-14  &	   1.80E-14   &    2.00E-14 & \\				
6548	[N II]	  &  7.39E-15  &     7.44E-15	&     8.35E-15  &	9.41E-15  &	  1.06E-14   &	   6.64E-15  &	   8.82E-15   &    7.29E-15 & \\
6563	H$\alpha$ &  5.61E-13  &     5.66E-13	&     5.24E-13  &	5.32E-13  &	  3.02E-13   &	   3.21E-13  &	   3.02E-13   &    2.91E-13 & \\  
6584	[N II]	  &  4.51E-14  &     3.96E-14	&     4.61E-14  &	4.71E-14  &	  3.73E-14   &	   2.90E-14  &	   3.46E-14   &    3.99E-14 & \\  
7065	He I	  &  4.69E-14  &     4.81E-14	&     4.76E-14  &	4.90E-14  &	  2.46E-14   &	   2.91E-14  &	   2.82E-14   &    2.21E-14 & \\
7135	[Ar III]  &  2.22E-14  &     2.10E-14	&     2.88E-14  &	3.02E-14  &	  2.10E-14   &	   1.69E-14  &	   2.74E-14   &    2.69E-14 & \\  
\hline
\multicolumn{4}{l}{Emission line peak observed wavelengths [ \AA ] } &  \\
4101	H$\delta$ & 4100.98  &   4100.97  &   4101.03 &  4100.97 & 4101.09 &  4101.05 &  4101.08 & 4100.94  & \\	   
4340	H$\gamma$ & 4339.66  &   4339.65  &   4339.64 &  4339.66 & 4339.78 &  4339.75 &  4339.67 & 4339.62  & \\ 
4363	[O III]   & 4362.38  &   4362.38  &   4362.37 &  4362.37 & 4362.46 &  4362.45 &  4362.36 & 4362.36  & \\
4861	H$\beta$  & 4860.44  &   4860.44  &   4860.43 &  4860.43 & 4860.54 &  4860.54 &  4860.40 & 4860.39  & \\  
4959	[O III]   & 4958.07  &   4958.07  &   4958.07 &  4958.06 & 4958.21 &  4958.21 &  4958.10 & 4958.10  & \\  
5007	[O III]	  & 5005.98  &   5005.99  &   5005.98 &  5005.97 & 5006.12 &  5006.12 &  5006.01 & 5006.00  & \\	
5876	He I	  & 5874.65  &   5874.63  &   5874.63 &  5874.63 & 5874.77 &  5874.76 &  5874.58 & 5874.65  & \\				
6548	[N II]	  & 6547.30  &   6547.33  &   6547.28 &  6547.29 & 6547.40 &  6547.42 &  6547.36 & 6547.30  & \\
6563	H$\alpha$ & 6561.72  &   6561.73  &   6561.71 &  6561.71 & 6561.87 &  6561.87 &  6561.61 & 6561.61  & \\  
6584	[N II]	  & 6582.58  &   6582.57  &   6582.56 &  6582.56 & 6582.67 &  6582.68 &  6582.52 & 6582.58  & \\  
7065	He I	  & 7063.86  &   7063.85  &   7063.81 &  7063.82 & 7064.01 &  7064.01 &  7063.91 & 7063.83  & \\
7135	[Ar III]  & 7134.67  &   7134.67  &   7134.61 &  7134.61 & 7134.90 &  7134.84 &  7134.72 & 7134.64  & \\  
\hline
\end{tabular}
\end{center}
\label{tab2}
\end{table*}
%%%--------------------------------------------------------------------------------------
\section{Tidal interaction between the components}

The physics of tidal interaction for stars with convective
envelopes has been analyzed by several authors.
We use the estimate from Zahn (1977,
1989). The synchronization timescale in terms of the period is
\begin{equation}
 \tau_{\rm syn} \approx 800 \left( \frac{ M_{\rm g}  R_{\rm g}}{ L_{\rm g}}\right)^{1/3} 
 \frac{M_{\rm g}^2 (\frac{M_{\rm g}}{M_{\rm wd}} + 1)^2}{R_{\rm g}^6} P_{\rm orb}^4\ \;   {\rm yr},
\label{sync}
\end{equation}
where $M_{\rm g}$ and $M_{\rm wd}$ are the masses of the giant and white dwarf
respectively in Solar units, and $R_{\rm g}$, $L_{\rm g}$ are the radius and
luminosity of the giant, also in Solar units. The orbital period $P_{\rm orb}$ is
measured in days. 

Following Hurley, Tout \& Pols (2002), 
the circularization time scale is:
\begin{equation}
 \frac{1}{{\tau}_{\rm circ}} = \frac{21}{2} \left(\frac{k}{T}\right) q{_2} (1+q{_2}) 
 \left(\frac{R_{\rm g}}{a}\right)^8 .
\label{circ}
\end{equation}
where the mass ratio $q_2 = M_{\rm wd}/M_{\rm g}$.
In  Eq.~\ref{circ}, $(k/T)$ is 
derived from Rasio et al. (1996), where $\it{k}$ is an apsidal motion constant and $\it{T}$ is the timescale on which significant
changes in the orbit take place through tidal evolution:
\begin{equation}
\left(\frac{k}{T}\right) = \frac{2}{21} \frac{f_{\rm conv}}{{\tau}_{\rm conv}} \frac{M_{\rm env}}{M_{\rm g}}  
\;  {\rm yr}^{-1} ,
\label{KT}
\end{equation}
where $M_{\rm env}$ is the envelope's mass, and
\begin{equation}
{\tau}_{\rm conv} = 0.4311 \left(\frac{M_{\rm env}R_{\rm env}(R_{\rm g}-\frac{1}{2}R_{\rm env})}{3 L_{\rm g}}\right) ^{\frac{1}{3}}  {\rm yr}
\label{tau}
\end{equation}
where $R_{\rm env}$ is the depth of the convective envelope. ${\tau}_{\rm conv}$ is the eddy turnover time scale (the time scale on which the largest convective cells turnover). 
The numerical factor $f_{\rm conv}$ is
\begin{equation}
{f}_{\rm conv} = {\rm min} \left[1, \left( \frac{P_{\rm tid}}{2{\tau}_{\rm conv}} \right) ^2 \right],
\label{fconv}
\end{equation}
where $P_{\rm tid}$ is the tidal pumping time scale given by
\begin{equation}
\frac{1}{P_{\rm tid}} = \left|\frac{1}{P_{\rm orb}} - \frac{1}{P_{\rm rot}}\right|.
\label{ptid}
\end{equation}
From Hut (1981) we can estimate the $\tau_{\rm al}$ to $\tau_{\rm circ}$ ratio:
\begin{equation}
\frac{\tau_{\rm al}}{\tau_{\rm circ}} = \frac{7}{\alpha + 3}.
\label{al}
\end{equation}
where $\tau_{\rm al}$ is alignment time scale and $\alpha$ is a dimensionless quantity, representing the ratio of the orbital and rotational angular
momentum:
\begin{equation}
\alpha = \frac {q_2}{1+q_2} \frac {1}{r_{\rm g}^{2}} \left( \frac {a}{R_{\rm g}} \right) ^2 ,
\end{equation}
where  $r_{\rm g}$ is the gyration radius of the giant. For stars with similar stellar parameters, the gyration radius is $r_{\rm g}$=0.14 (Claret 2007).

We assume for the red giant $R_{\rm env}=0.9~R_{\rm g}$ and $M_{\rm env}= 3.0~M_\odot$ (Herwig 2005).
Using stellar parameters $R_g$=75\rsun, $M_g$=6\msun, $L_g$=3160\lsun and $M_{wd}$=0.75\msun (Gromadzki et al. 2011), 
we calculate $P_{\rm tid}= 50$~days, 
$f_{\rm conv}=1$, $\tau_{\rm conv}=0.29$~yr, 
and $(k/T)= 0.055$ yr$^{-1}$. From Eq.~\ref{sync}, Eq.~\ref{circ} and Eq.~\ref{al} 
we calculate the tidal time scales.

If $P_{orb}=1700$~d, then the 
synchronization time scale $\tau_{\rm sync} = 5.75 \times 10^7$~yr; the 
circularization timescale $\tau_{\rm circ} = 1.4 \times 10^{10}$~yr; the
alignment timescale $\tau_{\rm al} = 6.5 \times 10^7$~yr; the
pseudo-synchronization time scale $\tau_{\rm ps} = 2.2 \times 10^7$~yr,
and the tidal force does not play an important role.  \\

If  $P_{orb}=245.68$~d, then      
$\tau_{\rm sync} = 2.5 \times 10^4$~yr, 
$\tau_{\rm circ} = 4.6 \times 10^5$~yr,
$\tau_{\rm al}   = 2.8 \times 10^4$~yr,
$\tau_{\rm ps}   = 1 \times 10^4$~yr, and the orbital eccentricity could be as high as 
$ e \approx 0.64$, if the system is pseudo-synchronized. 

Corradi \& Schwarz (1997) obtained 4000~yr for the age of the nebula around AS~201, and 
40000~yr for that around V417~Cen. This means that  the tidal force  plays important role only if the shorter photometric 
period ($\sim 250$~d) is the orbital one.

\section{Emission lines and spectral line variability in 2004}

\subsection{Diagnostic diagram $\lambda 5007/H\beta$ vs. $\lambda 4363/H\gamma$ }

The diagnostic diagram $\lambda 5007/H\beta$ vs. $\lambda 4363/H\gamma$ for 174 flux measurements 
(planetary nebulae, S,D,D'-type symbiotic stars, and 6 peculiar objects) is presented 
by Baella (2010).  

We calculate the ratio $R_1 = \lambda 5007/H\beta = 4.95, 4.47, 5.37, 5.19$  and  	
$R_2=\lambda 4363/H\gamma = 1.016,	0.916,	0.832,	1.154 $ for 
JD~2453107, JD~2453109, JD~2453159, JD~2453207 respectively. For every JD we use the average value from the two spectra
obtained in the night. This corresponds to $ 4.5 < R_1/R_2  < 6.4$.
However the symbiotic limit is  $R_1/R_2 \le 3 $ (see Fig.1 of Baella 2010). This means that V417~Cen is located above the symbiotic limit - somewhere between 
symbiotics and peculiar objects. 

%  What does it mean?????

\subsection{Density and temperature of the forbidden lines region}

In the Table~\ref{tab2} are listed the emission lines we observe in the spectrum. 
The critical densities of the forbidden lines, $N_{cr}$, are as follows: 
$\log N_{cr}=7.5$ for [O III]$\lambda$4363, $\log N_{cr}=5.8$ for [O III]$\lambda$5007, 
$\log N_{cr}=4.9$ for [N II]$\lambda$6584, $\log N_{cr}=6.7$ for [Ar III]$\lambda$7135 (Appenzeller \& Oestreicher, 1988). 
We do not detect emission lines with  $\log N_{cr} < 4.9$ (e.g. [S II]$\lambda$6730),
and we estimate the density of the forbidden lines region $\log N_e \approx 4.5 \pm 0.5$.

Using the latest collision strengths and line ratios for forbidden [O III]  lines (Palay et al. 2012, 
Nahar \& Ethan, private communication),
we obtain an  estimation of the temperature in the forbidden lines region  
in the range  $T_e= 75000 - 130000$~K (for $\log N_e \approx 4.5$).
It seems that there is a tendency for $T_e$ to decrease as the brightness increases.  
% we obtain a rough estimation of the density and the temperature in the forbidden lines region  
% $\log N_e \approx 4.2 \pm 0.4$ $T_e=120000 \pm 20000$~K.

\subsection{Emission lines variability}

The emission line spectrum of V417~Cen is strongly variable. 
In 1988 only [O~III] and He~I lines were seen in emission; in 1993 the 
Balmer lines and [N~II] were also pronounced (Van Winckel et al. 1994). 
In the low resolution spectra  taken in 1996 the intensity of [O~III] lines 
dramatically decreased and seemed  similar to H$\alpha$ (Munari \& Zwitter 2002).

The photometric variability of the object at the time of our spectroscopic observations is shown
in Fig.\ref{AAVSO}. The photometric data are taken from AAVSO.
During the time of our observations, the brightness of the star varied as follows: 
at MJD2453107-2453109  V$\approx$12.6-12.7, 
at MJD2453159 V$\approx$12.3 and 
at MJD2453207 V$\approx$11.4 mag. 
We had the chance to obtain spectra during the time of brightness changes.
The brightness variability has been accompanied by changes in the emission lines. 
The variability of the H$\alpha$ line is immediately apparent (see Fig.\ref{halpha} and Table.\ref{tab1}).
When the brightness of the star was 12.6-12.7 mag, the H$\alpha$ emission line had equivalent 
width EW$_{H\alpha}$~$\approx$~6.4$\pm$0.2~\AA\ and FWZI~$\approx$~290~\kms,
with the line extending at zero intensity from -228 to +56 \kms and FWHM~$\approx$~52~\kms. 
When the star was brighter, at $\approx$~11.3~mag, then EW$_{H\alpha}$~$\approx$~2.9~\AA\ 
and FWZI~$\approx$~130~\kms, 
extending from -126 to +2~\kms, but with FWHM remaining practically the same. 

The Balmer emission lines, appearing in the extended envelopes of symbiotic stars usually have an ordinary nebular profile
with typical FWHM~$\approx$~100-150~\kms, for example AG~Dra (Tomova \& Tomov 1999). 
FWHM increases to 200~\kms\ only during the active phase. The basic mechanism determining their 
width is turbulence in the gas. 
The central narrow emission of V417~Cen is very similar. 
The appearance and the variability of the narrow component is relatively 
common in symbiotic stars (see also Ikeda \& Tamura 2004 and references therein).

%In spite of the decrease of the equivalent widths of the emission lines, 
%the fluxes and FWHM of the observed Balmer lines ($H\alpha$, $H\beta$, $H\gamma$) do not change.
%The emitted flux in $H\beta$ decrease, while the brightness increased.
%the emitted flux in $H\alpha$ also.  
%In the same time the fluxes of the forbidden lines ([O~III]4363, 4959, 5007, [N~II] 6548, 6584, 
%[Ar~III] $\lambda$7135)  seems to remain  almost constant. ????

%Most probably this is due to dust obscuration. The dust does not obscure the emission lines region. 
%Our interpretation is the continuum of G2~Ib-II supergiant 
%was obscured by the dust (at the time of our sp1-sp4) and after it the dust absorption 
%decreased (at the time when our spectra sp7 and sp8 are taken). During the time of the light minimum the 
%the G9 supergiant spectrum is veiled by dust. 
%The  emergence of the continuum sources (giant and white dwarf) from the dust obscuration leads to 
%{\bf (i)} increase of the V brightness, 
%{\bf (ii)} decrease of the equivalent widths (the fluxes of the Balmer lines remain almost constant) and 
%{\bf (iii)} probably an increase of the ionization by the emerged hot source (white dwarf) which leads to
%the observed increase of the fluxes of the forbidden lines. 
%This scenario is schematic and should be considered as a first attempt to explain the observed 
%emission line variability. 
It seems that the optical fluxes of the Balmer lines decrease, while
the brightness increased. The fluxes of the forbidden lines ([O~III]~$\lambda$4363, ~$\lambda$4959, ~$\lambda$5007, [N~II]~$\lambda$6548, ~$\lambda$6584, 
[Ar~III] $\lambda$7135)  seem to remain  almost constant.
%1 However, bearing in  mind the uncertains in the flux calibration, these results are not realiable. 
The EWs of both Balmer and forbidden lines decreased, while the
brightness increased. The measured wavelengths of the emission lines are constant within the observational errors.
All fluxes and equivalent widths are measured by measuring the whole emission line profiles.

\section{Discussion}

The Balmer emission lines, appearing in the extended envelopes of symbiotic stars, usually have 
an ordinary nebular profile
with typical FWHM~$\approx$~100-150~\kms, for example AG~Dra (Tomova \& Tomov 1999). 
FWHM increases to 200~\kms\ only during the active phase. The basic mechanism determining their 
width is turbulence in the gas. 
The appearance and the variability of the narrow component is relatively 
common in symbiotic stars (see also Ikeda \& Tamura 2004 and the references therein).
The central narrow emission of V417~Cen is very similar.

When the star brightness increases, the EW of the line decreases. The FWHM and line fluxes of the Balmer lines
remain unchanged. Our interpretation is that the continuum of G2~Ib-II supergiant 
was obscured by dust and after it the dust absorption 
decreased. In other words, during the time of the light minimum the G9 supergiant  spectrum is veiled by dust. 

Rapid rotation is a common property of the cool components of D' SSs (Jorissen et al. 2005, Zamanov et al. 2006). 
This fast rotation is due to the mass transfer (spin accretion from the former AGB) and/or the tidal force  
(Soker 2002;  Ablimit \& L\"u  2012). 
Fast rotation in D'-type symbiotics  probably leads to the formation of a circumstellar disk 
as in the classical Be stars, where the fast rotation of the B star 
expels matter in the equatorial regions (see  Porter \& Rivinius 2003). 
The geometry and kinematics of the circumstellar environment in Be stars is best explained 
by a rotationally supported relatively thin disk with very little outflow. 
The central B star is a fast rotator with a commonly quoted mean value of about 
70\% - 80\%  of the critical velocity, i.e. the ratio $v_{rot}/v_{cr} \sim 0.7$. 
Regarding the ratio $v_{rot}/v_{cr} \sim 0.7$ the rotation  of the G2Ib-II component is similar
to the Be stars. This is estimated as 
$ v \, sin \, i = 75$ \kms, which is  71\% of the critical value and implies 
a short rotation period of the mass donor $P_{rot} \le 51$~d (Zamanov et al. 2006). 
By analogy with the  Be stars and Be/X-ray binaries, we expect a rotationally supported  disk around the supergiant in V417~Cen 
with little outflow.

The far-IR data (Kenyon, Fernandez-Castro \& Stencel 1988) for D'-type symbiotics 
indicate optically thick dust clouds, with temperature $T_{dust} \sim 100 - 400$~K  (Kenyon, Fernandez-Castro \& Stencel 1988).
The IR colours of V417~Cen reveal the presence of warmer dust more similar to D-type symbiotics.
The IR energy distribution peaks in the L band (3.79 $\mu m$), corresponding to a dust temperature of 800 ~K. 
The infrared excess is broad and cannot be fitted with a single black body, indicating 
temperature stratification in the dust (van Winckel et al. 1994).
Angeloni et al. (2007),  for another D'-type symbiotic 
HD330036, detected three dust shells, which are probably circumbinary. 

The IR energy distribution of V417~Cen is similar to the distribution in D-type symbiotics
with resolved bipolar nebulae like BI~Cru, He 2-104 and R~Aqr (Schwarz \& Corradi 1992), 
with a broad IR excess peaking in the L band. In these systems, an excretion disk is thought to 
provide equatorial density enhancement. 

Where is the dust located? There are three suggestions for this:   
\begin{itemize}
\item    The hot component lies outside the dust shell that enshrouds the 
giant companion  (as suggested for D-type symbiotics by Kenyon, Fernandez-Castro \& Stencel 1986).  
\item    The dust is located around the hot component and at $L_4$ and $L_5$  Lagrangian points (Gromadzki et al. 2011). 
\item    Dust shells  are circumbinary,  i.e. the orbital period is short and the supergiant and WD are in the centre of the dust shells, 
as supposed by  Angeloni et al. (2007) for HD330036.
\end{itemize}

\section*{Conclusion}
We report 8 high resolution spectra of the D'-type symbiotic star 
V417~Cen obtained between April - July 2004, when the star's V-band brightness increased from 12.4$^m$ to 11.5$^m$.

We find that the equivalent widths of the emission lines decreased, while the brightness increased.
The FWHMs and wavelengths of the emission lines do not change.

It seems that the fluxes of the observed Balmer lines ($H\alpha$, $H\beta$,
$H\gamma$) decreased by 20\%, while the brightness increased.
The fluxes of the forbidden lines ([O~III]$\lambda$4363, $\lambda$4959, $\lambda$5007, [N~II] $\lambda$6548, $\lambda$6584, 
[Ar~III] $\lambda$7135)  remained almost constant.

On the basis of the KI~$\lambda$7698.965 interstellar line, 
the interstellar extinction towards  V417~Cen is estimated as  $E_{B-V}$ = 0.95 $\pm0.10$.

Using the [O~III] lines, we obtain a rough estimation of the density and the temperature in the forbidden line region  
$\log N_e \approx 4.5 \pm 0.5$ and $T_e=100000 \pm 25000$~K.

\section*{Acknowledgments}  
We thank the anonymous referee for constructive comments.
This work was supported in part by the OP "HRD", ESF and
Bulgarian Ministry of Education and Science under the contract BG051PO001-3.3.06-0047. 
AG acknowledges funding from the Slovenian Research Agency and from the Centre of
Excellence for Space Sciences and Technologies SPACE-SI, an operation partly
financed by the European Union, European Regional Development Fund and Republic of
Slovenia, Ministry of Education, Science and Sport.

%\newpage


\begin{thebibliography}

\bibitem[2012]{AL} Ablimit, I., L\"u, G L., 2012, Sci. Chi.: Phys.Mech. Astron. 56, 663

\bibitem[2007]{AC} Angeloni, R., Contini, M., Ciroi, S., \& Rafanelli, P.\ 2007, A\&A, 472, 497 

\bibitem[Appenzeller \& Oestreicher(1988)]{1988AJ.....95...45A} Appenzeller, I., \& Oestreicher, R.\ 1988, AJ, 95, 45 

\bibitem[Baella(2010)]{2010IAUS..262..307B} Baella, N.~O.\ 2010, IAU  Symposium, 262, 307 

\bibitem[Cieslinski et al.(1994)]{1994A&AS..106..243C} Cieslinski, D., Elizalde, F., \& Steiner, J.~E.\ 1994,  A\&AS, 106, 243 

\bibitem[Claret(2007)]{2007A&A...467.1389C} Claret, A.\ 2007, A\&A 467, 1389 

\bibitem[Corradi et al.(2003)]{2003ASPC..303.....C} Corradi, R.~L.~M., 
Mikolajewska, J., \& Mahoney, T.~J.\ 2003, Symbiotic Stars Probing Stellar Evolution,
Astronomical Society of the  Pacific Conference Series, 303  

\bibitem[Corradi \& Schwarz(1997)]{1997ppsb.conf..147C} Corradi, R., \& Schwarz, H.~E.\ 1997, Physical Processes in Symbiotic Binaries and Related Systems, 147 

\bibitem[Frankowski \& Jorissen(2007)]{2007BaltA..16..104F} Frankowski, A., \& Jorissen, A.\ 2007, BaltA, 16, 104 

\bibitem[Gromadzki et al.(2011)]{2011apn5.confP..63G} Gromadzki, M., Mikolajewska, J., Pilecki, B., 
Whitelock, P., \& Feast, M.\ 2011, Asymmetric Planetary Nebulae 5 Conference, 
A. A. Zijlstra, F. Lykou, I. McDonald, and E. Lagadec, eds., 63P

\bibitem[Herwig(2005)]{2005ARA&A..43..435H} Herwig, F.\ 2005, ARA\&A, 43, 435 

\bibitem[Hurley et al.(2002)]{2002MNRAS.329..897H} Hurley, J.~R., Tout, C.~A., \& Pols, O.~R.\ 2002, MNRAS, 329, 897 

\bibitem[Hut(1981)]{1981A&A....99..126H} Hut, P.\ 1981, A\&A, 99, 126

\bibitem[Ikeda \& Tamura(2004)]{2004PASJ...56..353I} Ikeda, Y., \& Tamura, S., 2004, PASJ, 56, 353 

\bibitem[Jorissen et al.(2005)]{2005A&A...441.1135J} Jorissen, A., Za{\v c}s, L., Udry, S., Lindgren, H., \& Musaev, F.~A.\ 2005, A\&A, 441, 1135

\bibitem[Kaufer et al.(1999)]{1999Msngr..95....8K} Kaufer, A., Stahl, O., 
Tubbesing, S., Norregaard, P., Avila, G., Francois, P., Pasquini, L., \& 
Pizzella, A.\ 1999, The Messenger, 95, 8 

\bibitem[Kenyon et al.(1986)]{1986AJ.....92.1118K} Kenyon, S.~J., 
Fernandez-Castro, T., \& Stencel, R.~E.\ 1986, AJ, 92, 1118 

\bibitem[Kenyon et al.(1988)]{1988AJ.....95.1817K} Kenyon, S.~J., 
Fernandez-Castro, T., \& Stencel, R.~E.\ 1988, AJ, 95, 1817 

\bibitem[Munari \& Zwitter(1997)]{1997A&A...318..269M} Munari, U., \& Zwitter, T.\ 1997,  A\&A, 318, 269 

\bibitem[Munari \& Zwitter(2002)]{2002A&A...383..188M} Munari, U., \& Zwitter, T.\ 2002, A\&A, 383, 188

\bibitem[Palay et al.(2012)]{2012MNRAS.423L..35P} Palay, E., Nahar, S.~N., Pradhan, A.~K., \& Eissner, W.\ 2012, MNRAS, 423, L35 

\bibitem[Pereira et al.(2003)]{2003ASPC..303...85P} Pereira, C.~B., Cunha, K., 
\& Smith, V.~V.\ 2003, Symbiotic Stars Probing Stellar Evolution, 303, 85 

\bibitem[Porter \& Rivinius(2003)]{2003PASP..115.1153P} Porter, J.~M., \& Rivinius, T.\ 2003, PASP, 115, 1153 

\bibitem[Rasio et al.(1996)]{1996ApJ...470.1187R} Rasio, F.~A., Tout, C.~A., Lubow, S.~H., \& Livio, M.\ 1996, ApJ, 470, 1187 

\bibitem[Soker(2002)]{2002MNRAS.337.1038S} Soker, N.\ 2002, MNRAS, 337, 1038 

\bibitem[Steiner et al.(1988)]{1988ASPC....1...67S} Steiner, J.~E., Cieslinski, D., \& Jablonski, F.~J., 
1988, Progress and Opportunities in Southern Hemisphere Optical Astronomy.~ The CTIO 25th Anniversary Symposium, 1, 67 

\bibitem[Schwarz \& Corradi(1992)]{1992A&A...265L..37S} Schwarz, H.~E., \& Corradi, R.~L.~M.\ 1992, A\&A, 265, L37 

\bibitem[Tomova \& Tomov(1999)]{1999A&A...347..151T} Tomova, M.~T., \& Tomov, N.~A., 1999, A\&A, 347, 151 

\bibitem[Van Winckel et al.(1994)]{1994A&A...285..241V} Van Winckel, H., Schwarz, H.~E., Duerbeck, H.~W., \& Fuhrmann, B., 1994, A\&A, 285, 241 

\bibitem[Zahn(1977)]{1977A&A....57..383Z} Zahn, J.-P.\ 1977, A\&A, 57, 383 

\bibitem[Zahn(1989)]{1989A&A...220..112Z} Zahn, J.-P.\ 1989, A\&A, 220, 112 

\bibitem[Zamanov et al.(2006)]{2006MNRAS.365.1215Z} Zamanov, R.~K., Bode, M.~F., Melo, C.~H.~F., et al.\ 2006, MNRAS, 365, 1215 

\end{thebibliography}
\end{document}